\documentclass[11pt]{article}
\usepackage{typearea}\typearea{13}
\usepackage{amsmath}
\usepackage{amssymb}\usepackage{amsthm}
\usepackage{amsfonts}
\usepackage[dvipdfmx]{hyperref}\hypersetup{hidelinks}
\begin{document}
\theoremstyle{plain}
\newtheorem{thm}{Theorem}[section]
\newtheorem{lem}[thm]{Lemma}
\newtheorem{prop}[thm]{Proposition}
\newtheorem{cor}[thm]{Corollary}
\theoremstyle{definition}
\newtheorem{assum}[thm]{Assumption}
\newtheorem{notation}[thm]{Notation}
\newtheorem{defn}[thm]{Definition}
\newtheorem{clm}[thm]{Claim}
\newtheorem{ex}[thm]{Example}
\theoremstyle{remark}
\newtheorem{rem}[thm]{Remark}
\newcommand{\unit}{\mathbb I}
\newcommand{\ali}[1]{{\mathfrak A}_{[ #1 ,\infty)}}
\newcommand{\alm}[1]{{\mathfrak A}_{(-\infty, #1 ]}}
\newcommand{\nn}[1]{\lV #1 \rV}
\newcommand{\br}{{\mathbb R}}
\newcommand{\dm}{{\rm dom}\mu}
\newcommand{\lb}{l_{\bb}(n,n_0,k_R,k_L,\lal,\bbD,\bbG,Y)}
\newcommand{\Ad}{\mathop{\mathrm{Ad}}\nolimits}
\newcommand{\Proj}{\mathop{\mathrm{Proj}}\nolimits}
\newcommand{\RRe}{\mathop{\mathrm{Re}}\nolimits}
\newcommand{\RIm}{\mathop{\mathrm{Im}}\nolimits}
\newcommand{\Wo}{\mathop{\mathrm{Wo}}\nolimits}
\newcommand{\Prim}{\mathop{\mathrm{Prim}_1}\nolimits}
\newcommand{\Primz}{\mathop{\mathrm{Prim}}\nolimits}
\newcommand{\ClassA}{\mathop{\mathrm{ClassA}}\nolimits}
\newcommand{\Class}{\mathop{\mathrm{Class}}\nolimits}
\def\qed{{\unskip\nobreak\hfil\penalty50
\hskip2em\hbox{}\nobreak\hfil$\square$
\parfillskip=0pt \finalhyphendemerits=0\par}\medskip}
\def\proof{\trivlist \item[\hskip \labelsep{\bf Proof.\ }]}
\def\endproof{\null\hfill\qed\endtrivlist\noindent}
\def\proofof[#1]{\trivlist \item[\hskip \labelsep{\bf Proof of #1.\ }]}
\def\endproofof{\null\hfill\qed\endtrivlist\noindent}
\newcommand{\caA}{{\mathcal A}}
\newcommand{\caB}{{\mathcal B}}
\newcommand{\caC}{{\mathcal C}}
\newcommand{\caD}{{\mathcal D}}
\newcommand{\caE}{{\mathcal E}}
\newcommand{\caF}{{\mathcal F}}
\newcommand{\caG}{{\mathcal G}}
\newcommand{\caH}{{\mathcal H}}
\newcommand{\caI}{{\mathcal I}}
\newcommand{\caJ}{{\mathcal J}}
\newcommand{\caK}{{\mathcal K}}
\newcommand{\caL}{{\mathcal L}}
\newcommand{\caM}{{\mathcal M}}
\newcommand{\caN}{{\mathcal N}}
\newcommand{\caO}{{\mathcal O}}
\newcommand{\caP}{{\mathcal P}}
\newcommand{\caQ}{{\mathcal Q}}
\newcommand{\caR}{{\mathcal R}}
\newcommand{\caS}{{\mathcal S}}
\newcommand{\caT}{{\mathcal T}}
\newcommand{\caU}{{\mathcal U}}
\newcommand{\caV}{{\mathcal V}}
\newcommand{\caW}{{\mathcal W}}
\newcommand{\caX}{{\mathcal X}}
\newcommand{\caY}{{\mathcal Y}}
\newcommand{\caZ}{{\mathcal Z}}
\newcommand{\bbA}{{\mathbb A}}
\newcommand{\bbB}{{\mathbb B}}
\newcommand{\bbC}{{\mathbb C}}
\newcommand{\bbD}{{\mathbb D}}
\newcommand{\bbE}{{\mathbb E}}
\newcommand{\bbF}{{\mathbb F}}
\newcommand{\bbG}{{\mathbb G}}
\newcommand{\bbH}{{\mathbb H}}
\newcommand{\bbI}{{\mathbb I}}
\newcommand{\bbJ}{{\mathbb J}}
\newcommand{\bbK}{{\mathbb K}}
\newcommand{\bbL}{{\mathbb L}}
\newcommand{\bbM}{{\mathbb M}}
\newcommand{\bbN}{{\mathbb N}}
\newcommand{\bbO}{{\mathbb O}}
\newcommand{\bbP}{{\mathbb P}}
\newcommand{\bbQ}{{\mathbb Q}}
\newcommand{\bbR}{{\mathbb R}}
\newcommand{\bbS}{{\mathbb S}}
\newcommand{\bbT}{{\mathbb T}}
\newcommand{\bbU}{{\mathbb U}}
\newcommand{\bbV}{{\mathbb V}}
\newcommand{\bbW}{{\mathbb W}}
\newcommand{\bbX}{{\mathbb X}}
\newcommand{\bbY}{{\mathbb Y}}
\newcommand{\bbZ}{{\mathbb Z}}
\newcommand{\str}{^*}
\newcommand{\lv}{\left \vert}
\newcommand{\rv}{\right \vert}
\newcommand{\lV}{\left \Vert}
\newcommand{\rV}{\right \Vert}
\newcommand{\la}{\left \langle}
\newcommand{\ra}{\right \rangle}
\newcommand{\ltm}{\left \{}
\newcommand{\rtm}{\right \}}
\newcommand{\lcm}{\left [}
\newcommand{\rcm}{\right ]}
\newcommand{\ket}[1]{\lv #1 \ra}
\newcommand{\bra}[1]{\la #1 \rv}
\newcommand{\lmk}{\left (}
\newcommand{\rmk}{\right )}
\newcommand{\al}{{\mathcal A}}
\newcommand{\md}{M_d({\mathbb C})}
\newcommand{\id}{\mathop{\mathrm{id}}\nolimits}
\newcommand{\Tr}{\mathop{\mathrm{Tr}}\nolimits}
\newcommand{\Ran}{\mathop{\mathrm{Ran}}\nolimits}
\newcommand{\Ker}{\mathop{\mathrm{Ker}}\nolimits}
\newcommand{\spn}{\mathop{\mathrm{span}}\nolimits}
\newcommand{\Mat}{\mathop{\mathrm{M}}\nolimits}
\newcommand{\UT}{\mathop{\mathrm{UT}}\nolimits}
\newcommand{\DT}{\mathop{\mathrm{DT}}\nolimits}
\newcommand{\GL}{\mathop{\mathrm{GL}}\nolimits}
\newcommand{\spa}{\mathop{\mathrm{span}}\nolimits}
\newcommand{\supp}{\mathop{\mathrm{supp}}\nolimits}
\newcommand{\rank}{\mathop{\mathrm{rank}}\nolimits}
\newcommand{\idd}{\mathop{\mathrm{id}}\nolimits}
\newcommand{\ran}{\mathop{\mathrm{Ran}}\nolimits}
\newcommand{\dr}{ \mathop{\mathrm{d}_{{\mathbb R}^k}}\nolimits} 
\newcommand{\dc}{ \mathop{\mathrm{d}_{\cc}}\nolimits} \newcommand{\drr}{ \mathop{\mathrm{d}_{\rr}}\nolimits} 
\newcommand{\zin}{\mathbb{Z}}
\newcommand{\rr}{\mathbb{R}}
\newcommand{\cc}{\mathbb{C}}
\newcommand{\ww}{\mathbb{W}}
\newcommand{\nan}{\mathbb{N}}\newcommand{\bb}{\mathbb{B}}
\newcommand{\aaa}{\mathbb{A}}\newcommand{\ee}{\mathbb{E}}
\newcommand{\pp}{\mathbb{P}}
\newcommand{\wks}{\mathop{\mathrm{wk^*-}}\nolimits}
\newcommand{\he}{\hat {\mathbb E}}
\newcommand{\ikn}{{\caI}_{k,n}}
\newcommand{\mk}{{\Mat_k}}
\newcommand{\mnz}{\Mat_{n_0}}
\newcommand{\mn}{\Mat_{n}}
\newcommand{\mkk}{\Mat_{k_R+k_L+1}}
\newcommand{\mnzk}{\mnz\otimes \mkk}
\newcommand{\hbb}{H^{k,\bb}_{m,p,q}}
\newcommand{\gb}[1]{\Gamma^{(R)}_{#1,\bb}}
\newcommand{\cgv}[1]{\caG_{#1,\vv}}
\newcommand{\gv}[1]{\Gamma^{(R)}_{#1,\vv}}
\newcommand{\gvt}[1]{\Gamma^{(R)}_{#1,\vv(t)}}
\newcommand{\gbt}[1]{\Gamma^{(R)}_{#1,\bb(t)}}
\newcommand{\cgb}[1]{\caG_{#1,\bb}}
\newcommand{\cgbt}[1]{\caG_{#1,\bb(t)}}
\newcommand{\gvp}[1]{G_{#1,\vv}}
\newcommand{\gbp}[1]{G_{#1,\bb}}
\newcommand{\gbpt}[1]{G_{#1,\bb(t)}}
\newcommand{\Pbm}[1]{\Phi_{#1,\bb}}
\newcommand{\Pvm}[1]{\Phi_{#1,\bb}}
\newcommand{\mb}{m_{\bb}}
\newcommand{\E}[1]{\widehat{\mathbb{E}}^{(#1)}}
\newcommand{\lal}{{\boldsymbol\lambda}}
\newcommand{\rar}{{\boldsymbol r}}
\newcommand{\oo}{{\boldsymbol\omega}}
\newcommand{\vv}{{\boldsymbol v}}
\newcommand{\bbm}{{\boldsymbol m}}
\newcommand{\kl}[1]{{\mathcal K}_{#1}}
\newcommand{\wb}[1]{\widehat{B_{\mu^{(#1)}}}}
\newcommand{\ws}[1]{\widehat{\psi_{\mu^{(#1)}}}}
\newcommand{\wsn}[1]{\widehat{\psi_{\nu^{(#1)}}}}
\newcommand{\wv}[1]{\widehat{v_{\mu^{(#1)}}}}
\newcommand{\wbn}[1]{\widehat{B_{\nu^{(#1)}}}}
\newcommand{\wo}[1]{\widehat{\omega_{\mu^{(#1)}}}}
\newcommand{\dist}{\dc}
\newcommand{\hpu}{\hat P^{(n_0,k_R,k_L)}_R}
\newcommand{\hpd}{\hat P^{(n_0,k_R,k_L)}_L}
\newcommand{\pu}{ P^{(k_R,k_L)}_R}
\newcommand{\pd}{ P^{(k_R,k_L)}_L}
\newcommand{\puuz}{P_{R}^{(n_0-1,n_0-1)}\otimes P^{(k_R,k_L)}_R}
\newcommand{\pddz}{P_{L}^{(n_0-1,n_0-1)}\otimes P^{(k_R,k_L)}_L}
\newcommand{\puu}{\tilde P_R}
\newcommand{\pdd}{\tilde P_L}
\newcommand{\qu}[1]{ Q^{(k_R,k_L)}_{R, #1}}
\newcommand{\qd}[1]{ Q^{(k_R,k_L)}_{L,#1}}
\newcommand{\hqu}[1]{ \hat Q^{(n_0,k_R,k_L)}_{R, #1}}
\newcommand{\hqd}[1]{ \hat Q^{(n_0,k_R,k_L)}_{L,#1}}
\newcommand{\eij}[1] {E^{(k_R,k_L)}_{#1}}
\newcommand{\eijz}[1] {E^{(n_0-1,n_0-1)}_{#1}}
\newcommand{\heij}[1] {\hat E^{(k_R,k_L)}_{#1}}
\newcommand{\cn}{\mathop{\mathrm{CN}(n_0,k_R,k_L)}\nolimits}
\newcommand{\ghd}[1]{\mathop{\mathrm{GHL}(#1,n_0,k_R,k_L,\bbG)}\nolimits}
\newcommand{\ghu}[1]{\mathop{\mathrm{GHR}(#1,n_0,k_R,k_L,\bbD)}\nolimits}
\newcommand{\ghdb}[1]{\mathop{\mathrm{GHL}(#1,n_0,k_R,k_L,\bbG)}\nolimits}
\newcommand{\ghub}[1]{\mathop{\mathrm{GHR}(#1,n_0,k_R,k_L,\bbD)}\nolimits}
\newcommand{\hfu}[1]{{\mathfrak H}_{#1}^R}
\newcommand{\hfd}[1]{{\mathfrak H}_{#1}^L}
\newcommand{\hfui}[1]{{\mathfrak H}_{#1,1}^R}
\newcommand{\hfdi}[1]{{\mathfrak H}_{#1,1}^L}
\newcommand{\hfuz}[1]{{\mathfrak H}_{#1,0}^R}
\newcommand{\hfdz}[1]{{\mathfrak H}_{#1,0}^L}
\newcommand{\CN}{\overline{\hpd}\lmk\mnzk \rmk\overline{\hpu}}
\newcommand{\cnz}[1] {\chi_{#1}^{(n_0)}}
\newcommand{\eu}{\eta_{R}^{(k_R,k_L)}}
\newcommand{\ezu}{\eta_{R}^{(n_0-1,n_0-1)}}
\newcommand{\ed}{\eta_{L}^{(k_R,k_L)}}
\newcommand{\ezd}{\eta_{L}^{(n_0-1,n_0-1)}}
\newcommand{\fii}[1]{f_{#1}^{(k_R,k_L)}}
\newcommand{\fiir}[1]{f_{#1}^{(k_R,0)}}
\newcommand{\fiil}[1]{f_{#1}^{(0,k_L)}}
\newcommand{\fiz}[1]{f_{#1}^{(n_0-1,n_0-1)}}
\newcommand{\zeij}[1] {e_{#1}^{(n_0)}}
\newcommand{\CL}{\ClassA}
\newcommand{\CLn}{\Class_2(n,n_0,k_R,k_L)}
\newcommand{\braket}[2]{\langle#1,#2\rangle}
\newcommand{\abs}[1]{\left\vert#1\right\vert}
\newtheorem{nota}{Notation}[section]
\def\qed{{\unskip\nobreak\hfil\penalty50
\hskip2em\hbox{}\nobreak\hfil$\square$
\parfillskip=0pt \finalhyphendemerits=0\par}\medskip}
\def\proof{\trivlist \item[\hskip \labelsep{\bf Proof.\ }]}
\def\endproof{\null\hfill\qed\endtrivlist\noindent}
\def\proofof[#1]{\trivlist \item[\hskip \labelsep{\bf Proof of #1.\ }]}
\def\endproofof{\null\hfill\qed\endtrivlist\noindent}
\newcommand{\ZZ}{\bbZ_2\times\bbZ_2}
\newcommand{\SSS}{\mathcal{S}}
\newcommand{\cs}{s}
\newcommand{\ct}{t}
\newcommand{\hS}{S}
\newcommand{\hal}[1]{{\bf [Hal: #1]}}

\begin{flushright}
\footnotesize
\end{flushright}
\noindent
{\Large\bf Lieb-Schultz-Mattis type theorems for quantum spin chains\\ without continuous symmetry}
\par\bigskip

\renewcommand{\thefootnote}{\fnsymbol{footnote}}
\noindent
Yoshiko Ogata\footnote{%
Graduate School of Mathematical Sciencesm
The University of Tokyo, Komaba, Tokyo 153-8914, Japan.
Supported in part by
the Grants-in-Aid for
Scientific Research, JSPS.}
and Hal Tasaki\footnote{%
Department of Physics, Gakushuin University, Mejiro, Toshima-ku, 
Tokyo 171-8588, Japan.
}
\renewcommand{\thefootnote}{\arabic{footnote}}
\setcounter{footnote}{0}

\begin{quotation}
\small\noindent
{\bf Abstract:}
We prove that  a quantum spin chain with half-odd-integral spin cannot have a unique ground state with a gap, provided that the interaction is short ranged, translation invariant, and possesses time-reversal symmetry or $\ZZ$ symmetry (i.e., the symmetry with respect to the $\pi$ rotations of spins about the three orthogonal axes).
The proof is based on the deep analogy between the matrix product state formulation and the representation of the Cuntz algebra in the von Neumann algebra $\pi(\caA_{R})''$ constructed from the ground state restricted to the right half-infinite chain.
\end{quotation}

\section{Introduction}
\label{s:intro}
Quantum spin systems have been active topics of research in both theoretical and mathematical physics.
See, e.g., \cite{BR1,BR2,Sutherland,ZengChenZhouWenBOOK,TasakiBook}.
The Lieb-Schultz-Mattis theorem \cite{LSM} is one of mathematically rigorous results for quantum spin chains which have had strong and long lasting impact on physics research.
The theorem, in the form extended to infinite chains by Affleck and Lieb \cite{AL}, is a no-go theorem which states that the standard Heisenberg spin chain with half-odd integral spin can never have a unique ground state with a nonvanishing energy gap.
The theorem has been extended to a large class of one-dimensional quantum many-body systems \cite{OYA,YOA,TasakiLSM} and also to systems in higher dimensions \cite{O,H1,H2,NS}.
See \cite{AizenmanNachtergaele1994} for a similar theorem which applies to a different class of quantum spin chains.

In the original theorem \cite{LSM} and all the extensions mentioned above, an essential assumption is that the spin system has U(1) symmetry, i.e., the invariance with respect the uniform spin rotation about a single axis.
The theorems for one-dimensional systems are proved with a variational state constructed by applying a gradual nonuniform U(1) rotation to the ground state.
The U(1) invariance of the ground state guarantees that the variational state has low excitation energy.

Recently it has been argued that Lieb-Schultz-Mattis type no-go theorems should be valid for quantum many-body systems which do not necessarily posses continuous symmetry \cite{ChenGuWEn2011,PTAV,WPVZ,Watanabe2018}.
In particular it was conjectured that {\em a quantum spin chain with half-odd-integral spin cannot have a unique ground state with a gap, when the interaction is short ranged, translation invariant, and possesses time-reversal symmetry or $\ZZ$ symmetry}\/ (i.e., the symmetry with respect to the $\pi$ rotations of spins about the three orthogonal axes).
This conjecture was first stated by Chen, Gu, and Wen \cite{ChenGuWEn2011} as (small) parts of their general classification based on the analysis of fixed-point matrix product states.\footnote{%
The case with $\ZZ$ symmetry is treated implicitly in section~V.B.4, and the case with time-reversal symmetry is discussed in section~V.C.}
Later, Watanabe, Po, Vishwanath, and Zaletel \cite{WPVZ} presented a proof that applies to matrix product states (MPS).\footnote{%
They only discuss the case with time-reversal symmetry, but the case with $\ZZ$ symmetry may be treated similarly.
We note that not all of the results in \cite{WPVZ} are mathematically rigorous.
But we expect that the argument in Supporting Information of \cite{WPVZ} which involves [S1--9] can be made rigorous by using standard techniques in matrix product states, provided that one only considers quantum spin chains. 
See also section~8.3.3 of \cite{TasakiBook} for proofs for MPS.
}
They assumed that a unique gapped ground state can be represented as an injective MPS, and examined the projective representation of the symmetry group on the set of matrices defining the MPS.
They found that the assumed symmetry is inconsistent with possible projective representations, thus excluding the existence of an injective MPS.

In the present paper, we prove the above conjecture and establish Lieb-Mattis-Schultz type no-go theorems for quantum spin chains without continuous symmetry.
See Theorems~\ref{main1} and \ref{main2} for precise statements.

Our proof is based on the deep correspondence, which was developed in \cite{bjp,BJ,Matsui3,bjkw,Matsui1,Matsui2}, between MPS \cite{FNW} and the Cuntz algebra associated with the translation in the half-infinite chain.
More precisely one can define operators $\cs_\mu$ (with $\mu=-S,-S+1,\ldots,S-1,S$) in the von Neumann algebra $\pi(\caA_{R})''$ constructed from the ground state restricted to the right half-infinite chain, provided that the ground state is pure, translation invariant, and satisfies the condition called the split property.
A remarkable fact, which is essential for us, is that the operators $\cs_\mu$ may be regarded as infinite dimensional analogues of matrices that define a MPS.
Then one can basically repeat the argument for MPS and examine projective representations of the symmetry group on the set of operators $\{\cs_\mu\}$ to prove the theorems.

As far as we know, the strategy to focus on projective representations (on matrices for MPS) of symmetry groups appeared in physics literature in 2008, in the context of the characterization of symmetry protected topological phases \cite{Perez-Garcia2008,PollmannTurnerBergOshikawa2010}.\footnote{%
It is likely that the idea to make use of projective representations, at least implicitly, can be found in earlier works on quantum spin systems.
}
We should note however that, in 2001, Matsui developed a mathematical theory for quantum spin chains based on projective representations of group symmetry, not only for MPS but for more general pure states which satisfy the split property \cite{Matsui1}.\footnote{%
We also learned that the emergence of projective representations in MPS was already realized by Fannes, Nachtergaele, and Werner back in 1990's, although the work was not published \cite{N}.  (See also the remark at the end of section~1 of \cite{Matsui1}.)}
Matsui then proved an extended version of Lieb-Schultz-Mattis theorem for a class of quantum spin chains with continuous symmetry.
In this sense we can regard the present work as a direct generalization of Matsui's work to certain systems without continuous symmetry.

\section{Setting and main results}
We start by summarizing standard setup of quantum spin chains on the infinite chain \cite{BR1,BR2}.
Let $S$ be an element of $\frac 12 \bbN$ and let $\SSS=\{-S,-S+1,\ldots,S-1,S\}$.
We denote the algebra of $(2S+1)\times (2S+1)$ matrices by $\Mat_{2S+1}$.
We fix an orthonormal basis $\{\psi_\mu\}_{\mu\in\SSS}$ of $\cc^{2S+1}$, and 
set $e_{\mu,\nu}=\ket{\psi_\mu}\bra{\psi_\nu}$ for each $\mu,\nu\in\SSS$.
Let $\hS_1, \hS_2, \hS_3\in\Mat_{2S+1}$ be the standard spin operators specified by $(\hS_1)^2+(\hS_2)^2+(\hS_3)^2=S(S+1)$ and the commutation relations $[\hS_1,\hS_2]=i\hS_3$, $[\hS_2,\hS_3]=i\hS_1$, and $[\hS_3,\hS_1]=i\hS_2$.

We denote the set of all finite subsets in ${\bbZ}$ by ${\mathfrak S}_{\bbZ}$.
For each $z\in\bbZ$,  let $\caA_{\{z\}}$ be an isomorphic copy of $\Mat_{2S+1}$, and for any finite subset $\Lambda\subset\bbZ$, let $\caA_{\Lambda} = \otimes_{z\in\Lambda}\caA_{\{z\}}$, which is the local algebra of observables. 
For finite $\Lambda$, the algebra $\caA_{\Lambda} $ can be regarded as the set of all bounded operators acting on
the Hilbert space $\otimes_{z\in\Lambda}{\bbC}^{2S+1}$.
We use this identification freely.
If $\Lambda_1\subset\Lambda_2$, the algebra $\caA_{\Lambda_1}$ is naturally embedded in $\caA_{\Lambda_2}$ by tensoring its elements with the identity. 
The algebra $\caA_{R}$ (resp. $\caA_L$) representing the half-infinite chain
is given as the inductive limit of the algebras $\caA_{\Lambda}$ with $\Lambda\in{\mathfrak S}_{\bbZ}$, $\Lambda\subset[0,\infty)$
(resp. $\Lambda\subset(-\infty -1]$). 
The algebra $\caA$, representing the two sided infinite chain
is given as the inductive limit of the algebras $\caA_{\Lambda}$ with $\Lambda\in{\mathfrak S}_{\bbZ}$. 
Note that $\caA_{\Lambda}$ for $\Lambda\in {\mathfrak S}_{\bbZ}$,
and $\caA_R$ can be regarded naturally as subalgebras of
$\caA$.
Under this identification, for each $z\in \bbZ$,
we denote the 
spin operators in $\caA_{\{z\}}\subset\caA$ by $\hS_{1}^{(z)}, \hS_{2}^{(z)}, \hS_{3}^{(z)}$.
We denote the set of local observables by $\caA_{\rm loc}=\bigcup_{\Lambda\in{\mathfrak S}_\bbZ}\caA_{\Lambda}
$.
We denote by $\tau_x$ the automorphisms on $\caA$ representing the space translation by  $x\in\bbZ$.

A mathematical model of a quantum spin chain is fully specified by its interaction $\Phi$.
An interaction is a map $\Phi$ from 
${\mathfrak S}_{\bbZ}$ into ${\caA}_{\rm loc}$ such
that $\Phi(X) \in {\caA}_{X}$ 
and $\Phi(X) = \Phi(X)^*$
for $X \in {\mathfrak S}_{\bbZ}$. 
An interaction $\Phi$ is translation invariant if
$\Phi(X+x)=\tau_x(\Phi(X))$, 
for all $x\in{\mathbb Z}$ and $X\in  {\mathfrak S}_{\bbZ}$, and  is of finite range if there exists  $m\in {\mathbb N}$ such that
$\Phi(X)=0$ for $X$ with diameter larger than $m$.
For an interaction $\Phi$ and a finite set $\Lambda\in{\mathfrak S}_{\bbZ}$, we define the local Hamiltonian
as
\begin{equation}\label{GenHamiltonian}
\lmk H_{\Phi}\rmk_{\Lambda}:=\sum_{X\subset{\Lambda}}\Phi(X).
\end{equation}

Let $\Phi$ be a translation invariant finite range interaction. For each $A\in \caA$ and $t\in{\mathbb R}$, the limit
\begin{equation}
\alpha_t^{\Phi}(A)=\lim_{\Lambda\to\bbZ} e^{it(H_{\Phi})_{\Lambda}} Ae^{-it(H_{\Phi})_{\Lambda}}
\end{equation}
exists and defines a strongly continuous one parameter group of automorphisms $\alpha^\Phi$ on $\caA$. 
(See \cite{BR2}.)
We denote the generator of $C^* $-dynamics $\alpha^{\Phi}$ by $\delta_{\Phi}$.

A state $\omega$ on $\caA$ is called an \mbox{$\alpha^{\Phi}$-ground} state
if the inequality
$
-i\,\omega(A^*{\delta_{\Phi}}(A))\ge 0
$
holds
for any element $A$ in the domain $\caD({\delta_{\Phi}})$ of ${\delta_\Phi}$.
Let $\omega$ be an $\alpha^\Phi$-ground state, with the GNS triple $(\caH_\omega,\pi_\omega,\Omega_\omega)$.
Then there exists a unique positive operator $H_{\omega,\Phi}$ on $\caH_\omega$ such that
$e^{itH_{\omega,\Phi}}\pi(A)\Omega_\omega=\pi_\omega(\alpha_t^\Phi(A))\Omega_\omega$,
for all $A\in\caA$ and $t\in\mathbb R$.
We call this $H_{\omega,\Phi}$ the bulk Hamiltonian associated with $\omega$.
Note that $\Omega_\omega$ is an eigenvector of $H_{\omega,\Phi}$ with eigenvalue $0$.

The following definition clarifies what we mean by a model with a unique gapped ground state.
See \cite{BR2} for the general theory.
\begin{defn}
We say that a model with an interaction $\Phi$ has a unique gapped ground state if 
(i)~the $\alpha^\Phi$-ground state, which we denote as $\varphi$, is unique, and 
(ii)~there exists $\gamma>0$ such that
$\sigma(H_{\varphi,\Phi})\setminus\{0\}\subset [\gamma,\infty)$, where  $\sigma(H_{\varphi,\Phi})$ is the spectrum of $H_{\varphi,\Phi}$.
\end{defn}
Note that the uniqueness of $\varphi$ implies that 0 is a non-degenerate eigenvalue of $H_{\varphi,\Phi}$.

We now introduce two types of symmetry relevant to the present study.
Time-reversal is the unique  antilinear unital \mbox{$*$-automorphism} $\Xi$ on $\caA$ satisfying
\begin{align}
\Xi(S_j^{z}) = -S_j^{(z)},\quad j=1,2,3,\quad z\in\bbZ.
\end{align}
(See Appendix \ref{TR} for the existence of such an automorphism.)
For a state $\omega$ on $\caA$, its time-reversal $\hat \omega$ is
given by 
\begin{equation}
\hat \omega(A)
=\omega(\Xi(A^*)),\quad
A\in \caA.
\end{equation}
We say $\omega$ is time-reversal invariant if $\omega=\hat\omega$.

We also introduce the $\bbZ_2\times\bbZ_2$ symmetry.
Let $R_j$ with $j=1,2,3$ be the unique set of unital \mbox{$*$-automorphisms} on $\caA$ satisfying
\begin{equation}
R_j\bigl(\hS_k^{(z)}\bigr)  
=
\left\{
\begin{gathered}
\hS_k^{(z)},\quad j=k\\
-\hS_{k}^{(z)},\quad j\neq k.
\end{gathered} 
\right.,\quad \quad j,k=1,2,3,
\end{equation}
for each $z\in \bbZ$.
Note that $R_j$ is essentially the global spin rotation by $\pi$ about the $j$-axis.

An interaction $\Phi$ is said to be time-reversal invariant
if $\Xi(\Phi(X))=\Phi(X)$
for all $X\in {\mathfrak S}_\bbZ$.
If $\omega$ is an \mbox{$\alpha^{\Phi}$-ground} state of a translation and time-reversal invariant finite range interaction $\Phi$,
then
$\hat \omega$ and $\omega\circ\tau_x$ for any $x\in\bbZ$ are also \mbox{$\alpha^{\Phi}$-ground} states.
In particular, if $\omega$ is a unique \mbox{$\alpha^{\Phi}$-ground} state, it is pure, time-reversal invariant,  
and translation invariant.
Similarly, 
an interaction $\Phi$ is $\bbZ_2\times\bbZ_2$ invariant if
 $R_j(\Phi(X))=\Phi(X)$ for $j=1,2$ and
for all $X\in {\mathfrak S}_\bbZ$.
Note that $\Phi(X)$ automatically satisfies $R_3(\Phi(X))=\Phi(X)$ .
If $\omega$ is an \mbox{$\alpha^{\Phi}$-ground} state of a translation and $\bbZ_2\times\bbZ_2$ symmetry invariant finite range interaction $\Phi$,
then
$\omega\circ R_j$ for $j=1,2,3$ and $\omega\circ\tau_x$ for $x\in\bbZ$ are also \mbox{$\alpha^{\Phi}$-ground} states.
In particular, if $\omega$ is a unique \mbox{$\alpha^{\Phi}$-ground} state, it is pure,  $\ZZ$ invariant, and translation invariant.

We are ready to state our main theorems.
\begin{thm}\label{main1}
Suppose that $S$ is a half-odd-integer.
Then a model with a translation and time-reversal invariant finite range interaction $\Phi$ does not have a unique gapped ground state.
\end{thm}
\begin{thm}\label{main2}
Suppose that $S$ is a half-odd-integer.
Then a model with a translation and $\bbZ_2\times\bbZ_2$ invariant finite range interaction $\Phi$ does not have  a unique gapped ground state.
\end{thm}

The theorems are  proven in Sections~\ref{main1p} and \ref{main2p}.

A typical interaction $\Phi$ with both time-reversal and $\ZZ$ symmetry, to which either Theorem~\ref{main1} or \ref{main2} apply, is that of the XYZ model,
\begin{equation}
\Phi(\{x,y\})=\sum_{j=1}^3J_j(x-y)\,S^{(x)}_jS^{(y)}_j,
\end{equation}
with $J_j(z)\in\bbR$ such that $J_z(z)=0$ if $|z|\ge m$ for some $m\in\bbN$. 
Note that the original Lieb-Schultz-Mattis theorem \cite{LSM,AL} applies only to the case with $J_1(z)=J_2(z)$.
An interaction with time-reversal symmetry but not necessarily with $\ZZ$ symmetry is
\begin{equation}
\Phi(\{x,y\})=\sum_{j,j'=1}^3J_j(x-y)\,S^{(x)}_jS^{(y)}_{j'},
\end{equation}
and an interaction with $\ZZ$ symmetry but not with  time-reversal symmetry is
\begin{equation}
\Phi(\{x,y,z\})=J(x-z,y-z)\,S^{(x)}_1S^{(y)}_2S^{(z)}_3.
\end{equation}

\bigskip
\noindent
{\em Remarks:}
1. We in fact prove a more general statement that there can be no  translation invariant pure states with the split property which have either time-reversal or $\ZZ$ symmetry.
But the unique gapped ground states seem to be the only physically meaningful examples.

2. We can relax the assumption of translation invariance.
The only property we need for the theorems is that the translation on the spin chain induces a unital endomorphism on  $\pi(\caA_{R})''$.
(See Theorem~\ref{cun}.)
This means that the interactions need not to be strictly translation invariant.
It suffices if the interaction satisfies the following boundedness condition
\begin{equation}
\sup_{x\in\bbZ}\sum_{X\ni x}|X|^{-1}\lV \Phi(X)\rV <\infty.
\end{equation}
and the unique ground state is
quasi-equivalent to its space translation by an arbitrary amount.

3. The requirement that the interaction $\Phi$ has a finite range can be weakened to that of an exponential decay.

\section{The split property}
\label{s:split}
Before going into the proofs, we review an important notion known as the split property in operator algebra.
We in particular recall two results which are  essential for us.

We here give the following definition of the split property, which is most suitable for our purpose.
It corresponds to the standard definition \cite{dl} in our setting (see \cite{Matsui2}).
See Definition~1.1 of \cite{Matsui2} for a different characterization, which is more physically motivated.
\begin{defn}
Let $\varphi$ be a pure state on $\caA$. Let $\varphi_R$ be the ristriciton of
$\varphi$ to $\caA_R$, and $(\caH,\pi,\Omega)$ be the GNS triple of $\varphi_R$.
We say $\varphi$ satisfies the split property with respect to $\caA_L$ and $\caA_R$,
if the von Neumann algebra $\pi(\caA_{R})''$ is a type I factor.
\end{defn}
The important connection between this abstract notion and our physical problem is given by the following result due to Matsui, which is based on Hastings's area law theorem \cite{area}.
\begin{thm}[Corollary 3.2 of \cite{Matsui2}]\label{matsui}
Let $\varphi$ be a (not necessarily unique) pure $\alpha^\Phi$-ground state of a translation invariant finite range interaction $\Phi$, and denote by  $H_{\varphi,\Phi}$ the corresponding bulk Hamiltonian. 
Assume that $0$ is a non-degenerate eigenvalue of $H_{\varphi,\Phi}$ and
there exists $\gamma>0$ such that $\sigma(H_{\varphi,\Phi})\setminus\{0\}\subset [\gamma,\infty)$.
Then $\varphi$ satisfies the  split property with respect to $\caA_L$ and $\caA_R$.
\end{thm}
Recall that a type I factor is isomorphic to $B(\caK)$, the set of all bounded operators 
on a  Hilbert space $\caK$. 
See \cite{takesaki}.
It has been shown in \cite{arv}, on the other hand, that a unital endomorphism of
$B(\caK)$ gives a representation of the Cuntz algebra.
Since the translation to the right by unit distance in the spin chain induces an endomorphism on $\pi(\caA_{R})''$ when $\varphi$ is translation invariant,
the following MPS-like object shows up.

\begin{thm}[\cite{arv,bjp,BJ}, Proof of Proposition 3.5 of \cite{Matsui1} and Lemma 3.5 of \cite{Matsui3}]\label{cun}
Let $\varphi$ be a translation invariant pure split state on $\caA$.
Let $\varphi_R$ be the restriction of $\varphi$ to $\caA_R$, and $(\caH,\pi,\Omega)$ be the GNS triple of $\varphi_R$.
Then there exist operators $\cs_\mu\in \pi(\caA_R)''$ with $\mu\in\SSS$
satisfying the following:
\begin{align}
&\cs_{\mu}^*\cs_{\nu}=\delta_{\mu\nu}\unit,\label{s1}\\
&\sum_{\mu\in\SSS} \cs_{\mu} \pi_{}(A) \cs_{\mu}^*=\pi\circ\tau_{1}(A),\quad A\in \caA_R \label{st}.\\
&\pi\lmk e_{\mu\nu}\otimes\unit_{[1,\infty)}\rmk
=\cs_\mu \cs_\nu^*\quad
\text{for all} \quad \mu,\nu\in\SSS.\label{sr}
\end{align}
Here $ e_{\mu\nu}\otimes\unit_{[1,\infty)}$ indicates element $e_{\mu\nu}$ in $\caA_{\{0\}}=\Mat_{2S+1}$
embedded into $\caA_R$.
\end{thm}
This theorem enables us to extend the argument for MPS to our general setting.

\section{Proof of Theorem \ref{main1}}\label{main1p}
Suppose that $S$ is a half-odd-integer.
Let $\Phi$ be a translation and time-reversal invariant finite range interaction.
We assume that the model with $\Phi$ has a unique gapped ground state, and derive a contradiction.
By the uniqueness, $\varphi$ is pure, time-reversal invariant,  
and translation invariant.
We thus see from Theorem \ref{matsui} that $\varphi$ satisfies the split property.
Let $(\caH,\pi,\Omega)$ be the GNS triple of
$\varphi_R:=\left.\varphi\right\vert_{\caA_R}$.
Since $\varphi$ is a translation invariant pure split state on $\caA$,
Theorem \ref{cun} guarantees that there are operators
$\cs_\mu\in \pi(\caA_R)''$ with $\mu\in\SSS$ satisfying (\ref{s1}), (\ref{st}), and (\ref{sr}).

Since $\varphi$ is invariant under the time-reversal, we can construct the corresponding antilinear unital \mbox{$*$-automorphism} $\hat\Xi$ of $\pi(\caA_R)''$, which satisfies
\begin{equation}
\hat\Xi(\pi(A))=
\pi(\Xi(A)),\quad A\in \caA_R,
\end{equation}
and $\hat\Xi^2=\id$.
See Lemma \ref{jlem}.
We then set
\begin{equation}
\ct_\mu=(-1)^{S+\mu}\,\hat\Xi(\cs_{-\mu}),\quad \mu\in\SSS, 
\label{tdef}
\end{equation}
which definition should be compared with the action (\ref{theta}) of the time-reversal map onto the basis states.
By direct calculations using (\ref{s1}), (\ref{st}), (\ref{sr}), we can show that
\begin{align}
&\ct_\mu^* \ct_\nu=\delta_{\mu\nu}\unit,\label{ts1}\\
&\sum_{\mu\in\SSS}\ct_\mu\pi(A)\ct_\mu^*=\pi\circ\tau_1(A),\quad A\in\caA_R,\label{tt}\\
&\pi( e_{\mu\nu}\otimes\unit_{[1,\infty)})
=\ct_\mu \ct_\nu^*,\label{et}\\
&\ct_\mu\in \pi(\caA_R)''.\label{t4}
\end{align}
To check (\ref{et}), we use (\ref{xi}) from Appendix \ref{TR}.
Let us multiply both the relations  (\ref{st}) and (\ref{tt}) by $\cs_\nu^*$ from the left and by $\ct_\lambda$ from the right.
Using the properties (\ref{s1}) and (\ref{ts1}),
we obtain
\begin{equation}
\cs_\nu^*\ct_\lambda \pi(A)= \pi(A)\cs_\nu^*\ct_\lambda,\quad A\in\caA_R.
\end{equation}
This means that 
\begin{equation}
\cs_\nu^*\ct_\lambda \in \pi(\caA_R)'' \cap \pi(\caA_R)'=\bbC\unit,
\end{equation}
where we get the final identity because $\pi(\caA_R)''$ is a factor.
Hence there are constants $c_{\nu\lambda}\in \bbC$  such that 
$\cs_\nu^*\ct_\lambda =c_{\nu\lambda}\unit$.
Then we obtain
\begin{align}
\ct_\lambda
=\sum_\nu \cs_{\nu} \cs_{\nu}^*\ct_\lambda
=\sum_\nu c_{\nu\lambda}\cs_\nu
\end{align}
Substituting this to the right-hand side of (\ref{et}), we obtain
\begin{align}\pi(e_{\mu\nu}\otimes \unit_{[1,\infty)})=
\ct_\mu \ct_{\nu}^*
=\sum_{\alpha\beta}c_{\alpha\mu}\overline{c_{\beta\nu}}
\cs_\alpha \cs_\beta^*
=\sum_{\alpha\beta}
c_{\alpha\mu}\overline{c_{\beta\nu}}\pi(e_{\alpha\beta}\otimes \unit_{[1,\infty)})
\end{align}
Comparing the coefficients (because $\Mat_{2S+1}$ is simple), we obtain 
\begin{align}
c_{\alpha\mu}\overline{c_{\beta\nu}}=\delta_{\alpha\mu}\delta_{\beta\nu}.
\end{align}
From this, we conclude
that $c_{\nu\lambda}=\delta_{\nu\lambda}c$ with some $c\in\bbC$ such that $|c|=1$, and hence
\begin{equation}
\ct_\lambda=c\cs_\lambda,\quad \lambda\in\SSS.
\end{equation}
By the definition (\ref{tdef}) of $\ct_\lambda$, we obtain
\begin{align}\label{sm}
c\cs_\mu =(-1)^{S+\mu}\,\hat\Xi(\cs_{-\mu}),\quad \mu\in\SSS,
\end{align}
which only involves $\cs_\mu$.
Replacing $\mu$ by $-\mu$, we get
\begin{align}
c\cs_{-\mu} =(-1)^{S-\mu}\,\hat\Xi(\cs_{\mu}),\quad \mu\in\SSS.
\end{align}
Substituting this to (\ref{sm}), we finally find
\begin{align}
c\cs_\mu=(-1)^{2S} c\cs_{\mu}
=-c\cs_\mu,\quad \mu\in\SSS,
\end{align}
where we recalled that $2S$ is odd.
Since $|c|=1$, this implies $\cs_\mu=0$ for all $\mu\in\SSS$, but this contradicts to (\ref{sr}).

\section{Proof of Theorem \ref{main2}}\label{main2p}
Suppose that $S$ is a half-odd-integer.
Let $\Phi$ be a translation and $\ZZ$ invariant finite range interaction.
We assume that the model with $\Phi$ has a unique gapped ground state, and derive a contradiction.
By the uniqueness, $\varphi$ is pure, translation invariant, and also invariant under $R_1$, $R_2$, and $R_3$.
We thus see from Theorem \ref{matsui} that $\varphi$ satisfies the split property.
Again let $(\caH,\pi,\Omega)$ be the GNS triple of
$\varphi_R:=\left.\varphi\right\vert_{\caA_R}$.
Since $\varphi$ is a translation invariant pure split state on $\caA$,
Theorem \ref{cun} guarantees that there are operators
$\cs_\mu\in \pi(\caA_R)''$ with $\mu\in\SSS$ satisfying (\ref{s1}), (\ref{st}), and (\ref{sr}).

Since $\varphi$ is invariant under $R_1$, $R_2$, and $R_3$, we can construct, through the standard procedure, the corresponding automorphisms $\hat R_1$, $\hat R_2$, and $\hat R_3$ on $\pi(\caA_R)''$.
See, e.g., Chapter~2 of \cite{BR1}.
The automorphisms satisfy
\begin{equation}
\hat R_j(\pi(A))=
\pi(R_j(A)),\quad A\in \caA_R,\quad j=1,2, 3,
\end{equation}
and
\begin{align}\label{rr}
\hat R_j^2=\id,\quad \hat R_1\circ\hat R_2\circ \hat R_3=\id.
\end{align}

Let us define the matrices $V_j\in\Mat_{2S+1}(\bbC)$ with $j=1,2,3$, which represent the $\pi$ rotation of a single spin about the $j$-axis, by
\begin{align}
V_j:=e^{-i\pi \hS_j}.
\end{align}
As our $S$ is a half-odd-integer, we have
\begin{align}\label{key}
V_j^2=-1,\quad V_jV_k=-V_kV_j\ (j\neq k),\quad
V_1V_2V_3=-1.
\end{align}
Note that $R_j(A)=V_j^*AV_j$ for $A\in\caA_{\{0\}}$.
This in particular means that
\begin{align}
\hat R_j(
\pi(
e_{\mu\nu}\otimes \unit_{[1,\infty)}
)
)
=\pi(
V_j^*e_{\mu\nu}V_j\otimes \unit_{[1,\infty)}
).\label{Re}
\end{align}

For each $j=1,2,3$ and $\mu\in\SSS$, we define
\begin{equation}
\ct_\mu^{(j)}=
\sum_{\nu} \braket{\psi_\nu}{ V_j^* \psi_\mu}\hat{R}_j(\cs_\nu).
\end{equation}
By using the properties of $\cs_\mu$ and (\ref{Re}), we can show that $\ct_\mu^{(j)}$ ($\mu\in\SSS$) for each $j$ satisfies the relations (\ref{ts1}), (\ref{tt}), (\ref{et}), and (\ref{t4}).
We can then repeat the discussion in the previous section to conclude
\begin{equation}
\cs_\mu=c_j\ct_\mu^{(j)},\quad \mu\in\SSS,\quad j=1,2,3,
\end{equation}
with some constants $c_j\in\bbC$ such that $|c_j|=1$.
From the definition of $\ct_\mu$, we obtain the following three relations for $\cs_\mu$:
\begin{align}\label{scs}
\cs_\mu=c_j \sum_{\nu} \braket{\psi_\nu}{ V_j ^*\psi_\mu}\hat R_j(\cs_\nu),\quad
\mu\in\SSS,\quad j=1,2,3.
\end{align}
We will show that these conditions are inconsistent.
First, by substituting (\ref{scs}) into itself, one finds
\begin{equation}
\cs_\mu=c_j^2
\sum_{\nu} \braket{\psi_\nu}{ (V_j^*)^2\psi_\mu}\hat R_j^2(\cs_\nu)=-(c_j)^2\cs_\mu,
\end{equation}
where we used (\ref{rr}) and (\ref{key}).
This implies  $c_j=\pm i$ for $j=1,2,3$.
Next, we substitute (\ref{scs}) with $j=2$ into that with $j=1$, and then substitute (\ref{scs})
with $j=3$ into the resulting relation to get
\begin{align}
\cs_\mu&=c_1\sum_{\nu} \braket{\psi_\nu}{ V_1 ^*\psi_\mu}\,\hat R_1(\cs_\nu)
\notag\\
&=c_1c_2\sum_{\nu} \braket{\psi_\nu}{ (V_1 V_2)^*\psi_\mu}\,\hat R_1\circ\hat R_2(\cs_\nu)
\notag\\
&=c_1c_2c_3\sum_{\nu} \braket{\psi_\nu}{ (V_1 V_2V_3)^*\psi_\mu}\,\hat R_1\circ\hat R_2 \circ\hat R_3(\cs_\nu)
\notag\\
&=-c_1c_2c_3 \cs_\mu
\end{align}
where  we again used (\ref{rr}) and (\ref{key}).
Hence we obtain $c_1c_2c_3=-1$, which is impossible because
 $c_j=\pm i$ for $j=1,2,3$.

\section{Discussion}
We proved Lieb-Schultz-Mattis type no-go theorems for translation invariant quantum spin chains with half-odd-integer spin, which have either time-reversal or $\ZZ$ symmetry.\footnote{%
We are indebted to Haruki Watanabe for the material in the present section.
}
The reason that we have studied these two types of symmetry is that both of the symmetry groups yield nontrivial projective representations when acting on a single spin.
This means that Theorems~\ref{main1} and \ref{main2} are the simplest prototypes of a series of similar results, which may be proved by essentially the same methods, that hold for quantum many-body systems with much more complicated symmetry.
See, e.g., \cite{Watanabe2018} for discussion of such symmetry.

It may be useful to compare the new no-go theorems with the original Lieb-Schultz-Mattis theorem and its extensions.
We have already stressed that the continuous U(1) symmetry is essential for the original theorem, but that is not enough.
One also needs a condition on the ``filling factor'' which guarantees that the variational state constructed by nonuniform U(1) rotation is orthogonal to the ground state.
The new no-go theorems, in contrast, requires only the symmetry condition, but the symmetry group must allow a nontrivial projective representation.
Note that this condition excludes U(1) symmetry.
We thus conclude that the original Lieb-Schultz-Mattis theorem and the new no-go theorems are distinct statements based on essentially different mechanisms.\footnote{
A class of models with $\rm U(1)\times\bbZ_2$ symmetry can be covered by both the theorems.
}

In the present paper we made use of the close analogy between the matrices defining a MPS and the generators $\cs_\mu$ (with $\mu\in\SSS$) of the representation of the Cuntz algebra to prove concrete (and physically meaningful) statements about quantum spin chains.
It would be desirable if one could further pursue this strategy to prove theorems on one-dimensional quantum many-body systems.

\bigskip
{\small
It is a pleasure to thank Haruki Watanabe for valuable discussion which was essential for the present work, and Tohru Koma for useful discussion and comments.
We also thank Taku Matsui and Bruno Nachtergaele for useful comments.
The present work was supported by JSPS Grants-in-Aid for Scientific Research nos.~16K05171 (Y.O.) and 16H02211 (H.T.).
}

\appendix
\section{Antilinear operators}
In this appendix we summarize several general properties of antilinear operators.
These results can be proved by straightforwardly extending the corresponding proofs for linear operators.
\begin{lem}
Let $\caH$ be a Hilbert space and $A:\caH\to\caH$ be a bounded antilinear operator.
Then there is a unique bounded antilinear operator $A^*:\caH\to\caH$
such that
\begin{align}
\braket{Ax}{y}=\braket{A^*y}{x},\quad x,y\in \caH.
\end{align}
\end{lem}
For a unital $C^*$-algebra $\caA$ we say $\Xi:\caA\to\caA$ is a an antilinear unital $*$-homomorphism if
the following hold:
\begin{align}
&\Xi(\alpha A+\beta B)=\bar\alpha\Xi(A)+\bar\beta \Xi(B),\\
&\Xi(AB)=\Xi(A)\,\Xi(B),\\
&\Xi(A^*)=(\Xi (A))^*,\quad A,B\in\caA\\
&\Xi(\unit)=\unit.
\end{align}
If it is bijective, we call it an antilinear unital \mbox{$*$-automorphism}.
\begin{lem}\label{tb}
Let $\Xi:\caA\to\caA$ be a an antilinear unital $*$-homomorphism on a unital $C^*$-algebra $\caA$.
Then we have
\begin{equation}
\lV \Xi(A)\rV\le \lV A\rV,
\end{equation}
for all $A\in\caA$.
\end{lem}
\begin{lem}\label{jlem}
Let $\caA$ be a unital $C^*$-algebra and $\Xi$ an antilinear unital \mbox{$*$-automorphism} on $\caA$, with $\Xi^2=\unit$.
Let $\omega$ be a state on $\caA$ which is invariant under $\Xi$, i.e.,
\begin{equation}
\omega(\Xi(A^*))=\omega(A),\quad
A\in \caA.
\end{equation}
Let $(\caH, \pi,\Omega)$ be the GNS representation of $\omega$.
Then there is an antilinear operator $J:\caH\to\caH$ satisfying
\begin{gather}\label{jpr}
J^*=J,\quad J^2=\unit,\\
J\pi(A)J^*=\pi(\Xi(A)),\quad A\in\caA,\\
J\pi(\caA)''J^*= \pi(\caA)''.
\end{gather}
In particular, $\Xi$ is extended to
an antilinear automorphism $\hat \Xi$ on  the von Neumann algebra  $\pi(\caA)''$
by
\begin{align}
\hat \Xi(x)=J xJ^*,\quad x\in \pi(\caA)''.
\end{align}
\end{lem}

\section{Time-reversal automorphism}\label{TR}
In this section we define the time-reversal automorphism on $\caA$.
We start by defining time-reversal on finite systems.
We take the standard basis states for a single spin, which satisfy
$S_3\psi_\mu=\mu\psi_\mu$ and 
$(S_1\pm iS_2)\psi_\mu=\sqrt{S(S+1)-\mu(\mu\pm1)}\,\psi_{\mu\pm1}$ for $\mu\in\SSS$.
For each $I\in\mathfrak{S}_\bbZ$, we let $\theta_I$ be the antilinear unitary on $\bigotimes_{I}\bbC^{2S+1}$ such that
\begin{equation}
\theta_I\bigotimes_{x\in I}\psi^{(x)}_{\mu_x}
=\bigotimes_{x\in I}(-1)^{S-\mu_x}\,\psi^{(x)}_{-\mu_x},
\label{theta}
\end{equation}
where $\psi^{(x)}_{\mu}$ denotes the copy of $\psi_\mu$ for spin at $x\in I$.
We then define the antilinear unital \mbox{$*$-automorphism} $\Xi_I$ on $\caA_I$ by
\begin{equation}
\Xi_I(A)=\theta_I^* A\theta_I,\quad A\in\caA_I.
\end{equation}
One easily checks that 
\begin{align}\label{xi}
\Xi_I\lmk \bigotimes_{x\in I}e^{(x)}_{\mu_x,\nu_x}\rmk
=
\bigotimes_{x\in I}\lmk (-1)^{2S+\mu_x+\nu_x}\,e^{(x)}_{-\mu_x,-\nu_x}\rmk,
\end{align}
where $e^{(x)}_{\mu,\nu}$ is the copy of $e_{\mu,\nu}$ in $\caA_{\{x\}}$.
It is also not difficult to see that 
\begin{equation}
\Xi_I(S_j^{z}) = -S_j^{(z)},\quad j=1,2,3,\quad z\in I.
\label{XiS}
\end{equation}
See, e.g., \cite{TasakiBook}.
In fact $\Xi_I$ is the unique antilinear unital \mbox{$*$-automorphism} on $\caA_I$ which satisfies (\ref{XiS}).

For $I,J\in{\mathfrak S}_\bbZ$ such that $I\subset J$, we clearly have
\begin{equation}
\left.\Xi_{J}\right\vert_{I}=\Xi_I
\end{equation}
From this, we may define $\Xi_0:\caA_{\rm loc}\to \caA_{\rm loc}$ by
\begin{equation}
\Xi_0(A)=\Xi_{I}(A),\quad A\in \caA_{I}. 
\end{equation}
(Recall that we regard $\caA_I$ as a subalegbra of $\caA$.)
As we have $\lV \Xi_I\rV\le 1$ by Lemma \ref{tb}  and $\caA_{\rm loc}$ is dense in $\caA$,
$\Xi_0$ can be extended to an antilinear unital $*$-automorphism $\Xi$ on $\caA$ with $\Xi^2=\unit$.
We call this $\Xi$ the time-reversal automorphism.
From the definition, it is straightforward to check that $\Xi$ and the translation $\tau_x$ commute.

\end{document}